%% file: MainFile.tex
\documentclass[
    preprint,
    superscriptaddress,
    preprintnumbers,
    nofootinbib,
    showkeys,
    amsmath, amssymb,
    aps,
    pra,
]{revtex4-2}
\usepackage[utf8]{inputenc}
\usepackage[T1]{fontenc}
\usepackage{lipsum}

\input{configs}

\begin{document}


\title{ Nonlinear Regge trajectories of quarkonia from holography }

\author{Nelson R. F. Braga}
\email{braga@if.ufrj.br}
\affiliation{Instituto de Física, Universidade Federal do Rio de Janeiro, Caixa Postal 68528, RJ 21941-972, Brazil}

\author{Yan F. Ferreira}
\email{yancarloff@pos.if.ufrj.br}
\affiliation{Instituto de Física, Universidade Federal do Rio de Janeiro, Caixa Postal 68528, RJ 21941-972, Brazil}


\begin{abstract}

We propose a holographic model for quarkonia using the WKB approximation with the Langer correction to properly reproduce nonlinear Regge trajectories of the form $m_n^2 = \beta (n + c_0)^{2/3} + c_1$. This form is expected from previous studies involving the solution of Cornell potential for heavy quark-antiquark interactions using a model based on the quadratic form of the spinless Salpeter-type equation.
The model fits experimental masses with very good accuracy. The corresponding decay constants also show a reasonable agreement with the results obtained from experimental data.

\end{abstract}




\maketitle

\newpage


\vspace{.5\baselineskip}
\section{Introduction}
\label{sec: Introduction}

From semiclassical arguments, supported by the phenomenology of light mesons, it is expected that the squared masses of such particles should grow linearly with their radial excitation quantum number $n$ \cite{Irving:1977ea}. For heavy mesons, on the other hand, the authors of \cite{Chen:2018hnx}, by employing the Bohr-Sommerfeld quantization approach to a model based on the quadratic form of the spinless Salpeter-type equation (QSSE) with the Cornell potential, show that the squared masses of quarkonia should grow with $n^{2/3}$ when $n$ is large. The experimental masses of charmonium and bottomonium states are fitted in Ref.~\cite{Chen:2018hnx} with an equation of the form
\begin{gather}
    m_n^2 = β (n+c_0)^{2/3} + c_1,
    \label{eq: phenomenological Regge trajectory - intro}
\end{gather}
with a good agreement with experimental values, especially when compared to the linear fit (see Ref.~\cite{A:2023bxv} for a similar form for the mass spectra for quarkonium).

The AdS/CFT correspondence~\cite{Maldacena:1998im} motivated many different AdS/QCD holographic models with the purpose of describing hadronic properties. The general idea is to modify the gravitational background with the introduction of mass scales. Some of the first models that appeared are the hard-wall~\cite{Polchinski:2001tt}, that was used to calculate masses of glueballs~\cite{Boschi-Filho:2002wdj} and the soft-wall model, that produces linear Regge trajectories for light mesons~\cite{Karch:2006pv}. More recently, the tangent model was introduced in order to fit not only masses, but also the decay constants of heavy mesons. This framework was used in order to describe the thermal dissociation of such particles in a quark-gluon plasma with temperature, density, magnetic field, and angular momentum~\cite{Braga:2017bml, Braga:2018zlu, Braga:2018hjt, Braga:2019yeh,Braga:2023fac, Braga:2024heu}.

In this work, we use the WKB approximation, considering the Langer correction, to build an AdS/QCD model that reproduces a Regge trajectory of the form~\eqref{eq: phenomenological Regge trajectory - intro}. The model is shown to provide also decay constants with a reasonable agreement with the results obtained from experimental data.

The structure of the present article is as follows: In Sec.~\ref{sec: Holographic AdS models with a dilaton background} we review the general form of holographic models with a dilaton background and the particular case of the soft-wall. Then, in the following sections we show a sequence of steps aimed to reproduce a Regge trajectory of the form \eqref{eq: phenomenological Regge trajectory - intro}. As the first building block of the model, in Sec.~\ref{sec: Mass shift} we explain how to add a term of mass shift that will generate the parameter $c_1$. In Sec.~\ref{sec: Nonlinear Regge trajectories}, we make some remarks on the application of the WKB method in holography and develop a method to approximate the integral that appears in the WKB quantization condition for large values of masses. These remarks will be important through the following sections. Then we reproduce the exponent of $2/3$ using the WKB approach explained before.  At the end of this section we introduce a new parameter that will be responsible for reproducing the parameter $c_0$ in \eqref{eq: phenomenological Regge trajectory - intro}. In Sec.~\ref{sec: Decay Constants} we discuss the calculation of the decay constants and add the last building block of the model. Finally, in Secs.~\ref{sec: Numerical Results} and \ref{sec: Conclusions}, we present the numerical results and conclusions, respectively.


\vspace{.5\baselineskip}
\section{Holographic A\lowercase{d}S/QCD models with a dilaton background}
\label{sec: Holographic AdS models with a dilaton background}

\vspace{.5\baselineskip}
\subsection{A quick review}
\label{subsec: A quick review}

In the type of AdS/QCD models considered here, vector mesons are represented by an Abelian vector field $V_m$ in five dimensions described by the action
\begin{gather}
    S = -\frac{1}{4 g_5^2} \int ḏ^4x \int_{0}^{∞} ḏz\, \sqrt{-g}\, ē^{-ϕ(z)} F^{mn} F_{mn},
    \label{eq: action}
\end{gather}
where $F_{mn}$ is related to the field $V_m$ through $F_{mn} = ∂_m V_n - ∂_n V_m$ and the indices are contracted with the metric of an $\AdS_5$ space, given by
\begin{gather}
    ḏs^2 = \frac{R^2}{z^2} \PAR{-ḏt^2 + ḏ\xx^2 + ḏz^2},
    \label{eq: metric}
\end{gather}
where $R$ is the radius of the AdS and the function $ϕ(z)$ is referred to as the dilaton. In the soft-wall model,
\begin{gather}
    ϕ(z) = ϕ_{\SW}(z) \equiv κ^2 z^2.
    \label{eq: dilaton soft-wall}
\end{gather}
It is important to remark that the parameter $κ$ represents a smooth  infrared energy cutoff, that breaks conformal invariance and makes it possible to represent hadronic states. 

Taking a Fourier component of the field $V_m$ representing a particle at rest, we write $V_ m = ε_m v(z) ē^{-îωt}$, where $ε_m$ is a unitary vector representing the polarization of the field and $ω$ is the energy of the particle. Since the particle is at rest, we identify the energy of the particle with its mass $m$, so that $ω = m$. The equations of motion produced by the action \eqref{eq: action} in this case is
\begin{gather}
    ω^2 v(z) - \PAR[4]{ \frac{1}{z} + ϕ'(z) } v'(z) + v''(z) = 0.
    \label{eq: equation of motion for v}
\end{gather}
It is convenient to define the function $ψ$ as
\begin{gather}
    ψ(z) = \sqrt{\frac{R}{z}} ē^{-ϕ(z)/2} v(z),
\end{gather}
so that the equation of motion \eqref{eq: equation of motion for v} assumes the form of a Schrödinger-like equation with eigenvalues $m_n^2$:
\begin{gather}
    -ψ''(z) + V(z)ψ(z) = m_n^2 ψ(z),
    \label{eq: equation of motion for ψ}
\shortintertext{with}
    V(z) = \frac{3}{4z^2} + \frac{1}{2z}ϕ'(z) + \frac{1}{4} ϕ'(z)^2 - \frac{1}{2}ϕ''(z).
    \label{eq: Schrödinger-like potential}
\end{gather}
Eqs.~\eqref{eq: equation of motion for v} and \eqref{eq: equation of motion for ψ} are valid for any dilaton $\phi$. When a dilaton is given, the simple evaluation of the right-hand side of \eqref{eq: Schrödinger-like potential} gives the potential $V$. At this point there are two different approaches that one can follow. Either choose the form of the dilaton and calculate the corresponding function $ V(z) $ or follow the opposite procedure.

For the particular case of the soft-wall model, $ϕ$ is given by \eqref{eq: dilaton soft-wall}, so that
\begin{gather}
    V_{\SW}(z) = \frac{3}{4z^2} + κ^4 z^2.
    \label{eq: Schrödinger-like potential - soft-wall}
\end{gather}

The normalized\footnote{By norm we mean the conserved charge associated to the field $V_m$. Therefore, saying that the field is normalized is equivalent to saying that the value of the integral $\int (R/z) ē^{-ϕ(z)} |v(z)|^2 ḏz = \int |ψ(z)|^2 ḏz$ is 1.} solutions of \eqref{eq: equation of motion for ψ} in the soft-wall case are
\begin{gather}
    ψ_n(z) = \sqrt{\frac{2κ}{n+1}} (κ z)^{3/2} ē^{-κ^2 z^2/2} L_n^1(κ^2 z^2),
\end{gather}
with the masses of the states with radial excitation number $n$ given by
\begin{flalign}
    && m_n^2 = 4 κ^2 (n+1)
    && \llap{\text{($n = 0, 1, 2...$)}. \hspace{1.0em}}
    \label{eq: Regge trajectory - soft-wall}
\end{flalign}


\vspace{.5\baselineskip}
\subsection{Heavy mesons}
\label{subsec: Heavy mesons}

The soft-wall model provides Regge trajectories of the form $m_n^2 \sim n$. However, since this model has only one parameter, $κ$, the squared mass of the ground state, $m_0^2 = 4κ^2$, is not independent of the slope of the Regge trajectory $m_{n+1}^2 - m_n^2 = 4κ^2$. For light mesons, it is not a problem, but heavy mesons like charmonium or bottomonium do not respect this property. The main reason is that the quark masses are not negligible in the case of heavy mesons. One can incorporate this issue by considering a relation of the form $m_n^2 = a + b n$ or, with the appropriate parameters:
\begin{flalign}
    && m_n^2 = M_q^2 + 4 κ^2 (n+1)
    && \llap{\text{($n = 0, 1, 2...$)}. \hspace{1.0em}}
    \label{eq: Regge trajectory for VI}
\end{flalign}
In order to interpret the parameter $M_q$, note that the mass of a bound state of two quarks can be written as the sum of the masses of its constituents, $\tssum_q m_q$, plus the binding energy $E_n$ between them. Therefore,
\begin{gather}
    m_n^2 = M_q^2 + 4 κ^2 (n+1) = \PAR[3]{\!\tssum_q m_q + E_n}^{\!2}.
    \label{eq: interpreting the parameter Mq - 1}
\end{gather}
Now, consider that we could turn the binding energy off, making $E_n = 0$. In this case, the masses of all states should collapse into one single value independent of $n$: the sum of the masses of the constituent quarks. For the expression $M_q^2 + 4 κ^2 (n+1)$ not to depend on $n$, $κ$ must be zero. Therefore,
\begin{gather}
    \at[2]{M_q}_{κ=0} = \tssum_q m_q,
    \label{eq: interpreting the parameter Mq - 2}
\end{gather}
meaning that, when $κ = 0$, the parameter $M_q$ is the the sum of the mass of the constituent quarks. We cannot keep this assertion if $κ$ is not zero, but we can assume that $M_q$ and $\tssum_q m_q$ should still be related also in this case.

As a first step toward a holographic model for heavy mesons, we analyze in the next section how to find a new dilaton that is capable of reproducing the spectrum given by Eq.~\eqref{eq: Regge trajectory for VI}. The tangent model of Refs.~\cite{Braga:2017bml, Braga:2017oqw, Braga:2018zlu, Braga:2023fac, Braga:2024heu, Braga:2018hjt, Braga:2019yeh, Braga:2021fey}, attempts to shift the masses and does this with satisfactory success. In that model, a term of the form $Mz$ is added to the dilaton. This term helps with shifting the predicted masses, making them closer to experimental values. This method, however is not accurate. In particular, the masses of bottomonium and charmonium states present a considerable discrepancy with respect to the experimental data. The success of that model lies on the reasonable fit of the decay constants of charmonium and bottomonium states, which is necessary to describe the thermal properties of these particles. In the following section we show an improved way of shifting the masses based on Eq.~\eqref{eq: Regge trajectory for VI}, solving the so-called Schrödinger inverse problem. That means: we show how to find a potential that generates a given mass spectrum.



\vspace{.5\baselineskip}
\section{Incorporating the heavy quark masses in the holographic approach}
\label{sec: Mass shift}

The field equation for the soft-wall model has the form
\begin{gather}
    -ψ_n'' + V_{\SW} ψ_n = 4 κ^2 (n+1) ψ_n \,.
    \label{eq: equation of motion soft-wall}
\end{gather}
We want to find a dilaton $ϕ_{I}$ that leads to a spectrum of squared masses with the form $M_q^2 + 4 κ^2 (n+1)$, appropriate to heavy mesons, so that the corresponding field equation must have the form
\begin{gather}
    -ψ_n'' + V_{I} ψ_n = \PAR[2]{M_q^2 + 4 κ^2 (n+1)} ψ_n,
    \label{eq: equation of motion with shifted eigenvaues}
\end{gather}
where $V_{I}$ is the new Schrödinger-like potential generated by the this new dilaton $ϕ_{I}$.

Eqs.~\eqref{eq: equation of motion soft-wall} and \eqref{eq: equation of motion with shifted eigenvaues} have the same form. They are equal if one imposes
\begin{align}
    &V_{I}(z) = V_{\SW}(z) + M_q^2 = \frac{3}{4z^2} + M_q^2 + κ^4 z^2.
\end{align}
Using Eq.~\eqref{eq: Schrödinger-like potential} one finds the dilaton $ϕ_{I}$ that generates the potential above. The general solution is
\begin{gather}
    ϕ_{I}(z)
    = κ^2 z^2
      - 2 \ln\COL[3]{
        U(M_q^2/4κ^2, 0, κ^2z^2) + C_1\, L(-M_q^2/4κ^2, -1, κ^2 z^2)
      }
      + C_2,
\end{gather}
where $C_1$ and $C_2$ are constants, $U(a,b,x)$ is the Tricomi confluent hypergeometric function, and $L(n,α,x) = L_{n}^{(α)}(x)$ is the generalized Laguerre function.

This dilaton goes to a constant when $z$ goes to zero. According to the appendix of \cite{Braga:2024heu}, this constant can be incorporated into the $\AdS$ radius, making the value of $ϕ(0)$ arbitrary. For convenience, we choose $C_2$ such that $ϕ(0) = 0$. This can be done by using the relation\footnote{See Eq. 13.2.21 at~\cite{NIST:DLMF:hypergeometric}.} $U(M_q^2/4κ^2, 0, 0) =  Γ(1+M_q^2/4κ^2)^{-1} $. On the other hand, the Laguerre function satisfies\footnote{This is proven by Eqs. 18.9.13, 18.5.12 and 16.2.1 at~\cite{NIST:DLMF:hypergeometric}.} $ L(-M_q^2/4κ^2, -1, 0 ) = 0$. So, in order to have $ϕ(0) = 0$, one chooses $C_2 = -2 \ln\COL[2]{ Γ\PAR{1 + M_q^2/4κ^2} }$. This gives
\begin{gather}
    ϕ_{I}(z)
    = κ^2 z^2
      - 2 \ln\COL[3]{ Γ\PAR{1 + M_q^2/4κ^2} \PAR[3]{
         U(M_q^2/4κ^2, 0, κ^2z^2) + C_1\, L(-M_q^2/4κ^2, -1, κ^2 z^2)
         }
      }.
\end{gather}

For recovering the soft-wall model when $M_q$ is zero, we set $C_1 = 0$, which gives,
\begin{gather}
    ϕ_{I}(z)
    = κ^2 z^2
      - 2 \ln\COL[3]{
        Γ\PAR{1 + M_q^2/4κ^2} U(M_q^2/4κ^2, 0, κ^2 z^2)
      }.
    \label{eq: new dilaton without the tangent}
\end{gather} 
Now that we found a way to incorporate the heavy quark masses in the holographic approach, in the following section we discus how to incorporate the nonlinearity of the mass spectra.



\vspace{.5\baselineskip}
\section{Nonlinear Regge trajectories in the holographic approach}
\label{sec: Nonlinear Regge trajectories}

Considering a model based on the quadratic form of the spinless Salpeter-type equation with a Cornell potential, it was found in Ref.~\cite{Chen:2018hnx} that the radial Regge trajectories for quarkonia are not linear. For large $n$ they behave as $m_n^2 \sim n^{2/3}$. Masses of charmonium and bottomonium were then fitted to the expression:
\begin{gather}
    m_n^2 = β (n+c_0)^{2/3} + c_1.
    \label{eq: phenomenological Regge trajectory - 1}
\end{gather}
It was shown in \cite{Chen:2018hnx} that the mass spectra obtained from this model finds good agreement with experimental results, even for light mesons.

We show in this section how to use the WKB approximation to build a holographic model that reproduces the same Regge trajectory \eqref{eq: phenomenological Regge trajectory - 1}. Other approaches to nonlinear Regge trajectories in holography, some of them also using the WKB approximation, can be found in \cite{MartinContreras:2020cyg, MartinContreras:2021bis, MartinContreras:2025wnh}. In this paper, however, we will look not only to the leading order in large $n$ ($m_n^2 \simeq β n^{2/3}$ in our case), but, instead, we will use the Langer correction to properly obtain the terms $c_0$ and $c_1$ from \eqref{eq: phenomenological Regge trajectory - 1}.


\vspace{.5\baselineskip}
\subsection{WKB approximation}
\label{subsec: WKB approximation}

In the first-order WKB approximation with two turning points, the quantization condition for the equation
\begin{gather}
    -ψ''(z) + \PAR[2]{V(z) - m^2}ψ(z) = 0
    \label{eq: Schrödinger equation}
\shortintertext{is given by}
    \int_b^a \sqrt{-Q(z)}\, dz = \PAR[2]{n+1/2}\pi,
    \label{eq: wrong 1st order WKB quantization condition}
\end{gather}
where $Q(z) = V(z) - m^2$, $n$ is an integer, and the points $a$ and $b$ are, respectively, the largest and the smallest turning points, i.e., the values of $z$ for which $Q(z) = 0$.

However, because of the presence of the term $3/4z^2$ in the Schrödinger-like potential \eqref{eq: Schrödinger-like potential}, the WKB approximation is not directly applicable. The WKB procedure produces a solution for $z$ close to zero and another solution for $z$ close to the turning point $b$. The behaviors of these two parts should match in between.\footnote{Formally, the large $z$ behavior of the WKB solution at small $z$ should coincide with the small argument behavior of the solution close to the turning point $b$.} The derivation of equation Eq.~\eqref{eq: wrong 1st order WKB quantization condition} depends on this match, known as the connection condition. In the present case, however, these behaviors do not agree. Indeed, the $3/4z^2$ term is analogous to the angular momentum term $ℓ(ℓ+1)/r^2$ in the quantum two-body problem, where the same issue occurs. In order to make the WKB approximation applicable to the quantum two body problem, one introduces the Langer transformation \cite{Langer:19370415}, which modifies the radial variable, mapping the origin $r=0$ into $-∞$. The wave function is also transformed in order to keep \eqref{eq: Schrödinger equation} with the form of a Schrödinger equation. After the changes, the regularity condition at the origin $r=0$ becomes a normalization condition at $-∞$. In the notation of our problem, the Langer transformation reads
\begin{gather}
    \qquad
    z = ē^{x}/κ
    \qquad\qquad\text{and}\qquad\qquad
    ψ(z) = ē^{x/2} \tilde{ψ}(x),
    \label{eq: Langer transformation}
\end{gather}
where $κ$ is a parameter from $V$ with dimension of mass.

With this, Eq.~\eqref{eq: Schrödinger equation} becomes
\begin{gather}
    -\tilde{ψ}''(x) + \PAR{ \frac{1}{4} + \frac{ē^{2x}}{κ^2}V(ē^x/κ) - \frac{ē^{2x}}{κ^2}m^2} \tilde{ψ}(x) = 0.
    \label{eq: Schrödinger equation variable x}
\end{gather}

The quantization condition for the new problem is
\begin{gather}
    \int_{\ln(κb)}^{\ln(κa)} \sqrt{-\tilde{Q}(x)}\, dx = \PAR[2]{n+1/2}\pi,
    \label{eq: 1st order WKB quantization condition variable x}
\shortintertext{with}
    \tilde{Q}(x) = \frac{1}{4} + \frac{ē^{2x}}{κ^2} \PAR[2]{V(ē^x/κ) - m^2},
\end{gather}
and with ${\ln(κa)}$ and ${\ln(κb)}$ being the turning points of $\tilde{Q}(x)$. By reverting the change of coordinates defined in \eqref{eq: Langer transformation}, the quantization condition takes the same form of Eq.~\eqref{eq: wrong 1st order WKB quantization condition}, but now with
\begin{gather} 
    Q(z) = U(z) - m^2,
    \qquad\text{and}\qquad
    U(z) = V(z) + \frac{1}{4z^2}.
    \label{eq: corrected potential in the 1st order WKB quantization condition}
\end{gather}
Note that $a$ and $b$ are the turning points of $Q(z)$ in its new form \eqref{eq: corrected potential in the 1st order WKB quantization condition} and that the "corrected" potential $U$ differs from the potential $V$ by $1/4z^2$. This difference is known as the Langer correction.

As the integral in the left-hand side of \eqref{eq: wrong 1st order WKB quantization condition} may not be solvable with a closed form in general, it will be useful to expand it in powers of $1/m$, that is considered to be small in the WKB approach. For this reason, we will make use of the series expansion
\begin{align}
    \sqrt{1-x} = \sum_{j=0}^{∞} \frac{(-1/2)_j}{j!} x^j,
    \label{eq: series of rqrt(1-x)}
\end{align}
where   $(a)_j = Γ(a+j)/Γ(j)$ is the Pochhammer symbol, to write
\begin{align}
    \int_b^a \sqrt{-Q(z)}\, dz
    = \sum_{j=0}^{∞} \frac{(-1/2)_j}{j!}\, m\! \int_b^a \frac{U^j}{m^{2j}}\, dz.
\end{align}

As an example that will be useful in the following sections, let $α$ be a positive constant and consider a potential of the form
\begin{gather}
    V(z) = \frac{3}{4z^2} + κ^{2+α}z^α
    \label{eq: Schrodinger-like potential with free α}
    \quad\implies\quad
    U(z) = \frac{1}{z^2} + κ^{2+α}z^α.
\end{gather}
In this case,
\begin{align}
    \int_b^a \sqrt{-Q(z)}\, dz
    & = \sum_{j=0}^{∞}\sum_{k=0}^{j} \frac{(-1/2)_j}{j!} \binom{j}{k} \frac{1}{m^{2j-1}} \int_b^a \PAR{\frac{1}{z^2}}^{\!k}\PAR{κ^{2+α}z^α}^{j-k}\, dz \nonumber\\
    & = \sum_{j=0}^{∞}\sum_{k=0}^{j} \frac{(-1/2)_j}{j!} \binom{j}{k} \frac{κ^{(2+α)(j-k)}}{m^{2j-1}} \PAR[2]{F_{j,k}(a) - F_{j,k}(b)},
    \label{eq: calculating the WKB integral - 1}
\end{align}\\[-1.25\baselineskip]
where\\[-1.25\baselineskip]
\begin{gather}
    F_{j,k}(z)
    = \LCHA[7]\omatrix{\dfrac{z^{αj - (2+α)k + 1}}{α j - (2+α)k + 1},
                                    &\;\text{if } α j - (2+α)k + 1 != 0,
                    \\ \ln{κz},     &\;\text{if } α j - (2+α)k + 1  = 0 }
\end{gather}
is a primitive of  $z^{αj - (2+α)k}$. The turning points $a$ and $b$ are solutions of $Q(z) = U(z) - m^2 = 0$, so that there are powers of $1/m$ hidden in them. Indeed, the largest and smallest turning points can be expanded in powers of $1/m$, respectively, as
\begin{align}
    a \simeq \frac{1}{κ}\PAR[4]{\frac{m}{κ}}^{\!2/α}\COL[4]{1 - \frac{1}{α}{\PAR[4]{\frac{κ}{m}}}^{\!2+4/α}}
    \qquad\text{and}\qquad
    b \simeq \frac{1}{m}\COL[4]{1 + \frac{1}{2}{\PAR[4]{\frac{κ}{m}}}^{\!2+α}}.\qquad
    \label{eq: turning points - free alpha}
\end{align}
Given this, collecting the powers of $m$ in \eqref{eq: calculating the WKB integral - 1} is not easy in general, but we can find the leading orders by inspection, and keep the integral up the order $m^0$. The leading contribution from $F_{j,k}(a)$ can be found by setting $k=0$, which maximizes the exponent in $a^{αj-(2+α)k+1}$. Therefore, we write
\begin{multline}
    \hspace{4em}
    \at[4]{\sum_{j=0}^{∞} \frac{(-1/2)_j}{j!} \binom{j}{k} \frac{κ^{(2+α)(j-k)}}{m^{2j-1}} F_{j,k}(a)}_{k=0} \\
    \simeq \sum_{j=0}^{∞} \frac{(-1/2)_j}{j!} \frac{1}{αj+1} \PAR[4]{\frac{m}{κ}}^{\!1+2/α} + \calO(m^{-1-2/α}).
    \hspace{4em}
\end{multline}
To calculate the series, we substitute $\frac{1}{αj+1} = \int_0^1 x^{αj} ḏx$ and read Eq.~\eqref{eq: series of rqrt(1-x)} from the right to the left. This gives
\begin{multline}
    \at[4]{\sum_{j=0}^{∞} \frac{(-1/2)_j}{j!} \binom{j}{k} \frac{κ^{(2+α)(j-k)}}{m^{2j-1}} F_{j,k}(a)}_{k=0}
    \simeq \PAR[4]{\frac{m}{κ}}^{\!1+2/α} \int_0^1 \sum_{j=0}^{∞} \frac{(-1/2)_j}{j!} \PAR[2]{x^{α}}^j ḏx \\
    = \PAR[4]{\frac{m}{κ}}^{\!1+2/α} \int_0^1 \sqrt{1-x^{α}}\, ḏx
    = \frac{\Beta(1/α,3/2)}{α} \PAR[4]{\frac{m}{κ}}^{\!1+2/α},
    \label{eq: calculating the WKB integral - 2}
\end{multline}\\[-1\baselineskip]
since\\[-1\baselineskip]
\begin{gather}
    \int_0^1 \sqrt{1-x^{α}}\, ḏx
    = \frac{1}{α} \int_0^1 t^{1/α-1} (1-t)^{3/2-1}ḏt
    = \frac{\Beta(1/α,3/2)}{α},
\end{gather}
where $\Beta$ is the beta function.

The next contribution from $a$ comes from $k=1$, but $m^{1-2j}F_{j,1}(a)$ is already of an order smaller than $m^0$.

The first contribution from $b$ comes from setting $k=j$, which minimizes the exponent of $b^{αj-(2+α)k+1}$. These terms give
\begin{align}
    &-\at[4]{\sum_{j=0}^{∞} \frac{(-1/2)_j}{j!} \binom{j}{k} \frac{κ^{(2+α)(j-k)}}{m^{2j-1}} F_{j,k}(b)}_{k=j} 
    \simeq \sum_{j=0}^{∞} \frac{(-1/2)_j}{j!} \frac{1}{2j-1} m^0 + \calO(m^{-2-α}).\quad
\end{align}
Now we separate the $j=0$ term and substitute
\begin{flalign}
    && \frac{1}{2j-1} = \int_0^1 x^{2j-2} ḏx
    && \llap{\text{($j!=0$)}~\hspace{1.0em}}
\end{flalign}
to obtain
\begin{multline}
    -\at[4]{\sum_{j=0}^{∞} \frac{(-1/2)_j}{j!} \binom{j}{k} \frac{κ^{(2+α)(j-k)}}{m^{2j-1}} F_{j,k}(b)}_{k=j}
    \simeq -1 + \int_0^1 \frac{1}{x^2}\PAR[4]{\sum_{j=0}^{∞} \frac{(-1/2)_j}{j!} \PAR[2]{x^{2}}^{\!j} - 1} ḏx \\
    = -1 + \int_0^1 \frac{1}{x^2}\PAR[3]{\sqrt{1-x^2} - 1} ḏx
    = -\frac{π}{2}.
    \label{eq: calculating the WKB integral - 3}
\end{multline}

The next order in $b$ comes from setting $k=j-1$, but it is clearly smaller than $m^0$.

Using the results \eqref{eq: calculating the WKB integral - 1}, \eqref{eq: calculating the WKB integral - 2} and \eqref{eq: calculating the WKB integral - 3}, one approximates the quantization condition as
\begin{gather}
    \int_b^a \sqrt{-Q(z)}\,ḏz
    \simeq \frac{\Beta(1/α,3/2)}{α} \PAR[4]{\frac{m}{κ}}^{\!1+2/α} - \frac{π}{2}
    = \PAR{n+\frac{1}{2}}π.
\end{gather}
Therefore,\vspace{-.5\baselineskip}
\begin{flalign}
    && m_n = \COL{ \frac{α π}{\Beta(1/α,3/2)} (n+1) }^{α/(2+α)} κ
       \hspace{5em}
    && \llap{\text{($n = 0, 1, 2...$)}. \hspace{1.0em}}
    \label{eq: Regge trajectory - free alpha}
\end{flalign}

As a consistency check for the procedure, note from Eq.~\eqref{eq: Schrodinger-like potential with free α} that when $α=2$, we recover the soft-wall model \eqref{eq: Schrödinger-like potential - soft-wall}. And, indeed, Eq.~\eqref{eq: Regge trajectory - free alpha} agrees with the exact soft-wall result found in Eq.~\eqref{eq: Regge trajectory - soft-wall} when $α=2$.


\vspace{.5\baselineskip}
\subsection{Regge trajectories with power 2/3 }
\label{subsec: 2/3 Regge Trajectories}

From Eq.~\eqref{eq: Regge trajectory - free alpha}, one notes that a Schrödinger-like potential of the form $V(z) = 3/4z^2 + κ^{2+α}z^{α}$ produces a Regge trajectory that goes as $m \sim (n+1)^{α/(2+α)}$. So, a trajectory of the form $m^2 \sim (n+1)^{2/3}$ is obtained by choosing $α = 1$. This gives
\begin{gather}
    V_{\romanII}(z) = \frac{3}{4z^2} + κ^3 z
\shortintertext{and}
    m_n^2 \simeq \PAR{\frac{3π}{2}}^{\!2/3}(n+1)^{2/3} κ^2.
\end{gather}
At this point we have the correct large $n$ behavior for the trajectory, but not yet the shift of masses in the first state of heavy mesons due to the heavy quark masses. In Sec.~\ref{sec: Mass shift} we  showed how to obtain a shift in the masses. We just have to add a term $M_q^2$ to the Schrödinger-like potential. So, we change the potential to 
\begin{gather}
    V_{\romanIII}(z) = \frac{3}{4z^2} + κ^3 z + M_q^2,
\shortintertext{so that}
    m_n^2 \simeq\PAR{\frac{3π}{2}}^{\!2/3}(n+1)^{2/3} κ^2 + M_q^2.
\end{gather}


\vspace{.5\baselineskip}
\subsection{Adjusting the parameter $c_0$}
\label{subsec: adjusting the c_0 parameter}

Recall that our objective is to obtain a Regge trajectory of the form
\begin{gather}
    m_n^2 = β (n+c_0)^{2/3} + c_1.
    \label{eq: phenomenological Regge trajectory - 2}
\end{gather}
At this point, we have $β = (3π/2)^{2/3}κ^2$, $c_1 = M_q^2$ and $c_0$ is the fixed and universal parameter $c_0 = 1$. However, the value $c_0 = 1$ does not agree with its value from the model in \cite{Chen:2018hnx}. The final step in order to build a model that yields $m_n^2 = β (n+c_0)^{2/3} + c_1$ will be to add a new term to the Schrödinger-like potential, thus promoting $c_0$ to an adjustable parameter.

The term that does this in the potential must be of the form $1/\sqrt{z}$, so that the potential becomes
\begin{gather}
    V_{\romanIV}(z) = \frac{3}{4z^2} + \frac{A κ^2}{\sqrt{κz}} + M_q^2 + κ^3 z,
\end{gather}
where $A$ is a dimensionless constant.

We will set $M_q$ to zero and recover it later by summing it to the squared masses at the end, shifting the masses of the model with $M_q = 0$, as we saw in Sec.~\ref{sec: Mass shift}.

Now we must update the WKB calculations of Sec.~\ref{subsec: WKB approximation}. The integral that appears in the quantization condition now is
\begin{align}
    \int_b^a \sqrt{-Q(z)}\, dz
    & = \sum_{j=0}^{∞}{\sum}' \frac{(-1/2)_j}{k_1!\, k_2!\, k_3!} \frac{1}{m^{2j-1}} \int_b^a \PAR[4]{\frac{1}{z^2}}^{\!\!k_3} \PAR[4]{\frac{A κ^{3/2}}{\sqrt{z}}}^{\!\!k_2} \PAR{κ^3 z}^{k_1}\, dz \nonumber\\
    & = \sum_{j=0}^{∞}{\sum}' \frac{(-1/2)_j}{k_1!\, k_2!\, k_3!} \frac{A^{k_2} κ^{k_1+3k_2/2}}{m^{2j-1}} \PAR[2]{F_{j;k_1,k_2,k_3}(a) - F_{j;k_1,k_2,k_3}(b)},
    \label{eq: calculating the WKB integral for VIII - 1}
\end{align}\\[-1.25\baselineskip]
where\\[-1.25\baselineskip]
\begin{gather}
    F_{j;k_1,k_2,k_3}(z)
    = \LCHA[7]\omatrix{\dfrac{z^{-2k_3-k_2/2+k_1+1}}{-2k_3-k_2/2+k_1+1},
                                    &\;\text{if } -2k_3-k_2/2+k_1+1 != 0,
                    \\ \ln{κz},     &\;\text{if } -2k_3-k_2/2+k_1+1  = 0. }
\end{gather}
We used the multinomial theorem, so that the summation with the prime is performed over $k_1$, $k_2$ and $k_3$ ranging from $0$ to $j$ and with the constraint  $k_1 + k_2 + k_3 = j$. The expressions for the turning points change to
\begin{align}
    \qquad
    a \simeq \frac{m^2}{κ^3} - \frac{A}{m}
    \qquad\quad\text{and}\qquad\quad
    b \simeq \frac{1}{m} + \frac{A κ^{3/2}}{2 m^{7/2}}.\qquad
\end{align}

The first contribution from $a$ to the integral \eqref{eq: quantization condition integral for VIV} comes from setting $k_1 = j$ and $k_2 = k_3 = 0$. Its calculation is very similar to what was done in \eqref{eq: calculating the WKB integral - 2}, but the the order $1/m$ in the expression for $a$ now leads to a contribution of order $m^0$ in $m^{1-2j}F_{j;j,0,0}(a)$. This smaller contribution, however, vanishes in the final summation. The result is
\begin{multline}
    \hspace{4em}
    \at[4]{\sum_{j=0}^{∞}{\sum}' \frac{(-1/2)_j}{k_1!\, k_2!\, k_3!} \frac{A^{k_2} κ^{k_1+3k_2/2}}{m^{2j-1}} F_{j;k_1,k_2,k_3}(a)}_{k_1=j,k_2=k_3=0} \\
    \simeq \sum_{j=0}^{∞} \frac{(-1/2)_j}{j!} \PAR{ \frac{1}{j+1}\frac{m^3}{κ^3} - A m^0 }
    = \frac{2}{3}\frac{m^3}{κ^3}.
    \hspace{4em}
\end{multline}

A new contribution from $a$ is now present, it comes from $k_1 = j-1$, $k_2 = 1$ and $k_3 = 0$:
\begin{multline}
    \hspace{4em}
    \at[4]{\sum_{j=0}^{∞}{\sum}' \frac{(-1/2)_j}{k_1!\, k_2!\, k_3!} \frac{A^{k_2} κ^{k_1+3k_2/2}}{m^{2j-1}} F_{j;k_1,k_2,k_3}(a)}_{k_1=j-1, k_2=1, k_3=0} \\
    \simeq \sum_{j=1}^{∞} \frac{(-1/2)_j}{(j-1)!} \frac{2A}{2j-1}\frac{m^3}{κ^3}
    = -\frac{Aπ}{2}.
    \hspace{4em}
    \label{eq: quantization condition integral for VIV}
\end{multline}

The contribution from $b$ is $-π/2$, exactly the same as the one found in Eq.~\eqref{eq: calculating the WKB integral - 3}. Then, we write the quantization condition as
\begin{gather}
    \int_b^a \sqrt{-Q(z)}\,ḏz
    \simeq \frac{2}{3} \frac{m^3}{κ^3} - \frac{Aπ}{2} - \frac{π}{2}
    = \PAR{n+\frac{1}{2}}π.
\end{gather}
Therefore, recovering the term $M_q^2$, the first order WKB approximation for the spectrum of squared masses is
\begin{flalign}
    && m_n^2 = \COL[4]{\frac{3π}{2} \PAR{ n + 1 + A/2 }}^{2/3} κ^2 + M_q^2
       \hspace{5em}
    && \llap{\text{($n = 0, 1, 2...$)}. \hspace{1.0em}}
\end{flalign}
Hence, in order to reproduce the Regge trajectory of Eq.~\eqref{eq: phenomenological Regge trajectory - 2} one chooses: 
\begin{gather}
    β   = (3π/2)^{2/3}κ^2,
    \qquad\quad
    c_0 = 1 + A/2,
    \qquad\text{and}\qquad
    c_1 = M_q^2.
    \label{eq: parameters of the Regge trajectory - holographic model}
\end{gather}



\vspace{.5\baselineskip}
\subsection{Comparison with the QSSE model}
\label{sec: Comparison to the QSSE model}

In Ref.~\cite{Chen:2018hnx} one finds an expression for the parameter $β$ from Eq.~\eqref{eq: phenomenological Regge trajectory - 2}. However, there are no expressions for the parameters $c_0$ and $c_1$. From the WKB approach with the Langer correction and with the method of collecting the leading orders in powers of $1/m$ explained here, one can write equations analogous to \eqref{eq: calculating the WKB integral - 1} and \eqref{eq: calculating the WKB integral for VIII - 1} for the QSSE model, and find expressions for the QSSE versions of the parameters $β$, $c_0$, and $c_1$ as well. This procedure gives
\begin{flalign}
    && m_n^2 = \COL[3]{12 π m_q σ \PAR{ n + 3/4 }}^{2/3} + 4 m_q^2
       \hspace{5em}
    && \llap{\text{($n = 0, 1, 2...$)}, \hspace{1.0em}}
    \label{eq: Regge trajectory - QSSE model}
\end{flalign}
so that we identify
\begin{gather}
    β   = (12 π m_q σ)^{2/3},
    \qquad\quad
    c_0 = 3/4,
    \qquad\quad\text{and}\qquad\quad
    c_1 = 4 m_q^2,
    \label{eq: parameters of the Regge trajectory - QSSE}
\end{gather}
where $m_q$ is the constituent quark mass and $σ$ is the string tension found in the linear term of the Cornell potential.

Comparison of \eqref{eq: parameters of the Regge trajectory - holographic model} with \eqref{eq: parameters of the Regge trajectory - QSSE} allows us to interpret our holographic parameters $κ$ and $M_q^2$. This confirms that $M_q$ can, indeed, be interpreted as the sum of the masses of the constituent quarks, since
\begin{gather}
    c_1 = M_q^2 = 4m_q^2 \implies M_q = 2m_q.
    \label{eq: parameter interpretation - 1}
\end{gather}
Additionally, we see that $κ$ is related to the string tension through
\begin{gather}
    β = (12 π m_q σ)^{2/3} = (3π/2)^{2/3}κ^2 \implies κ = (8 m_q σ)^{1/3}.
    \label{eq: parameter interpretation - 2}
\end{gather}
Finally, for the large $m$ behavior of our model to perfectly match the one given by the Cornell potential in the QSSE, we set $A = -1/2$, so that
\begin{gather}
    c_0 = 1 + A/2 = 3/4.
\end{gather}



\vspace{.5\baselineskip}
\section{Decay Constants}
\label{sec: Decay Constants}

In the type of holographic models considered here, decay constants of meson states are calculated from the formula \cite{Braga:2017bml}
\begin{gather}
    f_n = \frac{1}{g_5 m_n} e^{-\phi(0)} \lim_{z
\rightarrow 0} \frac{R}{z} v'(\omega,z),
\end{gather}
where $v$ is the field representing the meson and the constant
\begin{gather}
    g_5^{} = \sqrt{\frac{12 \pi^2}{N_c} R\, e^{-\phi(0)}} = 2 \pi \sqrt{R}\, e^{-\phi(0)/2},
\end{gather}
is fixed by imposing that the holographic propagator associated with the field $v$ matches the perturbative QCD result \cite{Grigoryan:2007my}.

The authors of \cite{Grigoryan:2010pj} pointed out that by strengthening the well in the the potential-like function $V$, one is capable of fitting charmonium's decay constants. In their model, they introduced some parameters to their potential in order to achieve this. Here, we found that adding a term of the form $ - B κ^2/√{κ z} $ to the small $z$ behavior of the potential, one can find good results of decay constants for both charmonium and bottomonium. To make this term to affect only the small $z$ behavior of the potential, we multiplied it by $ē^{-κ z}$, so that its influence at large $z$ values decreases exponentially. The new and final form of the potential is
\begin{gather}
    V(z) = \frac{3}{4z^2} - \frac{1}{2}\frac{κ^2}{\sqrt{κz}} - \frac{B κ^2}{\sqrt{κz}}ē^{-κz} + M_q^2 + κ^3 z.
    \label{eq: Shrodinger-like potential - final model}
\end{gather}

The dilaton that produces this potential is the solution of \eqref{eq: Schrödinger-like potential}, with $V$ given by \eqref{eq: Shrodinger-like potential - final model}. This time, however, the solution to \eqref{eq: Schrödinger-like potential} is not analytic, but we can find its large and small $z$ behaviors, which are
\begin{align}
    ϕ (z)
    &\simeq -2\ln\COL[3]{ C_1 \sqrt{z} \Ai\!\PAR[3]{κ z + M_q^2/κ^2} + C_2 \sqrt{z} \Bi\!\PAR[3]{κ z + M_q^2/κ^2} }
\end{align}
for $z$ large, and
\begin{gather}
         ϕ(z) \simeq -2 \ln\CHA{ C_3 \COL{\, 1 + \frac{4}{3} \PAR{B-\frac{1}{2}}(κz)^{3/2} \,} } 
\end{gather}
for small $z$. The symbols $\Ai$ and $\Bi$ represent the Airy functions.

We will choose the solution with $C_2 = 0$. This way we avoid problems, that also appear in the soft-wall model with the negative sign for the dilaton, such as the appearance of a null mass state. The large $z$ behavior of the dilaton is, then
\begin{gather}
    ϕ (z) \simeq \frac{4}{3}(κz)^{3/2} + \frac{2 M_q^2}{κ^2}\sqrt{κz} - \frac{1}{2}\ln\!\PAR{κz},
    \label{eq: dilaton for large z - model}
\end{gather}
where we expanded the Airy function for large values of its argument.

As in Sec.~\ref{sec: Mass shift}, we can set $ϕ(0) = 0$ for convenience. We do this by choosing $C_3 = 1$, so that 
\begin{gather}
        ϕ(z) \simeq -\frac{8}{3}\PAR{B-\frac{1}{2}}(κ z)^{3/2} 
    \label{eq: dilaton for samall z - model}
\end{gather}
for small $z$.

This analysis allows us to find the dilaton by numerically solving Eq.~\eqref{eq: Schrödinger-like potential} using the fact that the dilaton should match the large and small $z$ behaviors of Eqs.~\eqref{eq: dilaton for large z - model} and \eqref{eq: dilaton for samall z - model} as boundary conditions.

\vspace{.5\baselineskip}
\section{Numerical Results}
\label{sec: Numerical Results}

In this section we show the results found for the best fits and the values of the normalized root mean square error (NRMSE). In the literature, one finds this metric defined as
\begin{gather}
    \textrm{NRMSE}
    = 100\%
      \sqrt{ \dfrac{1}{N-N_p}
             \sqrtsum{i\,=\,1}{N} \PAR{\frac{y_i-\hat{y}_i}{\mean(\hat{y}_i)}}^{\!2} },
\end{gather}
where $N$ is the number of experimental points, $N_p$ is the number of parameters, the $y_i$'s are the predicted values of the quantity fitted, and the $\hat{y}_i$'s are its measured values.

Here, however, we are fitting two different quantities. For this reason, to calculate the errors for the combined sets of masses and decay constants of quarkonia, we define the combined normalized root mean square error (CNRMSE) as
\begin{gather} \label{eq: CNRMSE}
    \textrm{CNRMSE}
    = 100\%
      \sqrt{ \dfrac{1}{N-N_p} \COL[4]{
        \sqrtsum{n\,=\,0}{N_m-1}
          \PAR[4]{\frac{m_n-\hat{m}_n}{\mean(\hat{m}_n)}}^{\!\!2}
        + \sqrtsum{n\,=\,0}{N_{\!f}-1}
          \PAR[4]{\frac{f_n-\hat{f}_n}{\mean(\hat{f}_n)}}^{\!\!2}
             }},
\end{gather}
where the $m$'s are the masses, the $f$'s are the decay constants, $N_m$ is the number of measured masses, $N_{\!f}$ is the number of measured decay constants, $N = N_m + N_{\!f}$ is the total number of experimental points and, as before, $N_p = 3$ is the number of parameters.

Note that the presence of the number of parameters $N_p$ in the denominator imposes a penalty to a model with too many parameters, as compared to the number of data points.  For example, if a model with $N_p$ parameters is used to fit a set of $N=N_p$ experimental points, the value of the NRMSE would be infinite, which would suggest that this hypothetical model should be rejected.

All the experimental values were taken from the Particle Data Group (PDG) \cite{ParticleDataGroup:2020ssz} and we followed \cite{Kher:2018wtv} to use $ψ(4660)$ as the $5S$ state of the vector charmonium, but we do not have an experimental value for its decay constant.


\vspace{.5\baselineskip}
\subsection{Best fit parameters}
\label{subsec: Numerical Results - Best Fit Parameters}

The parameters that provided the best fits of masses and decay constants for charmonium and bottomonium are presented in Table~\ref{tab: best fits}. The bootstrap method was used in order to find the value of the parameters and their uncertainties. We created a sample of 2000 sets of masses and decay constants for each particle assuming that the distribution of each experimental value reproduce a Gaussian distribution with average being the expected value from PDG~\cite{ParticleDataGroup:2020ssz} and with standard deviation identified with the PDG's uncertainties.\footnote{For charmonium, each set is made of 9 experimental points: 5 masses and 4 decay constants; while for bottomonium, each set consists of 12 points: 6 masses and 6 decay constants.} We then fitted each of these 2000 sets of masses and decay constants. This gave us a distribution of values for the three parameter $κ$, $M_q$ and $B$. Finally, we identified the value of $κ$ and its uncertainty with, respectively, the average and standard deviation of values of best-fitting $κ$'s. The same was done for $M_q$ and $B$.

The value of the CNRMSE found for the best fitting parameters are $14.8\%$ for charmonium and $11.9\%$ for bottomonium. We also calculated the errors of masses and decay constants considered separately. We obtained $3\%$ for charmonium's masses, $17.7\%$ for charmonium's decay constants, $2.3\%$ for bottomonium masses, and $11.8\%$ for bottomonium decay constants.

\begin{table}[t]
\centering
\begin{tabular}{ccc}
    \rule{7em}{0pt} & \rule{11em}{0pt} & \rule{11em}{0pt} \\\hline\hline
    \multicolumn{3}{c}{\xrowht{3.5ex}Charmonium and bottomonium parameters for the best fits} \\\hline\hline\xrowht{5.2ex}
    Parameter &
    \makecell{\xrowht{5.2ex}\makecell{Charmonium\\values (\si{\giga\electronvolt})}} &
    \makecell{\xrowht{5.2ex}\makecell{Bottomonium\\values (\si{\giga\electronvolt})}}
    \\\hline\hline
    $κ$   & $ 1.161 \pm 0.008$ & $  2.109 \pm 0.061$ \\\hline
    $M_q$ & $ 3.12  \pm 0.04 $ & $  9.24  \pm 0.39 $ \\\hline
    $B$   & $ 7.42  \pm 0.15 $ & $ 11.20  \pm 0.47 $ \,\\\hline\hline
\end{tabular}\\[10pt]
\caption{Parameters that give the best fit of charmonium and bottomonium masses and decay constants.}
\label{tab: best fits}
\end{table}

\vspace{.5\baselineskip}
\subsection{Masses and decay constants}
\label{subsec: Numerical Results - Masses and Decay Constants}

\begin{figure}[htb]
    \centering
    \includegraphics{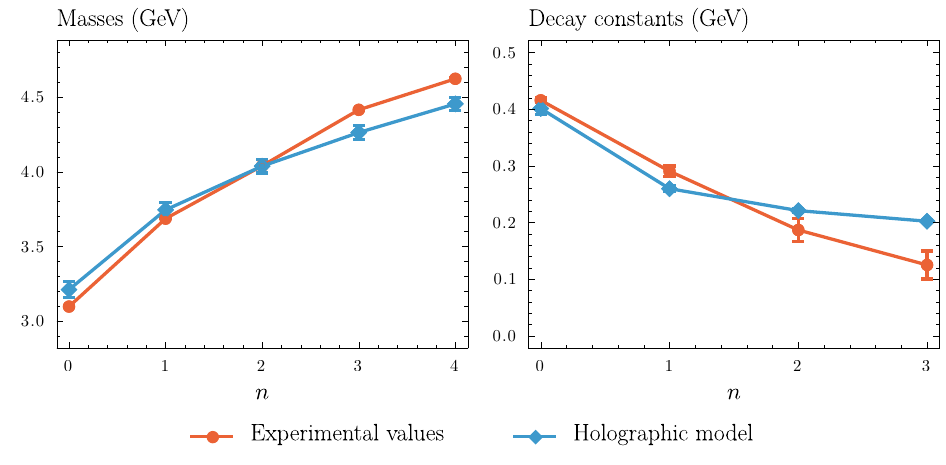}
    \caption{Comparison of charmonium experimental masses and decay constants with the ones given by the best fit of the holographic model.}
    \label{fig: charmonium masses and decay constants}
\end{figure}

\begin{figure}[htb]
    \centering
    \includegraphics{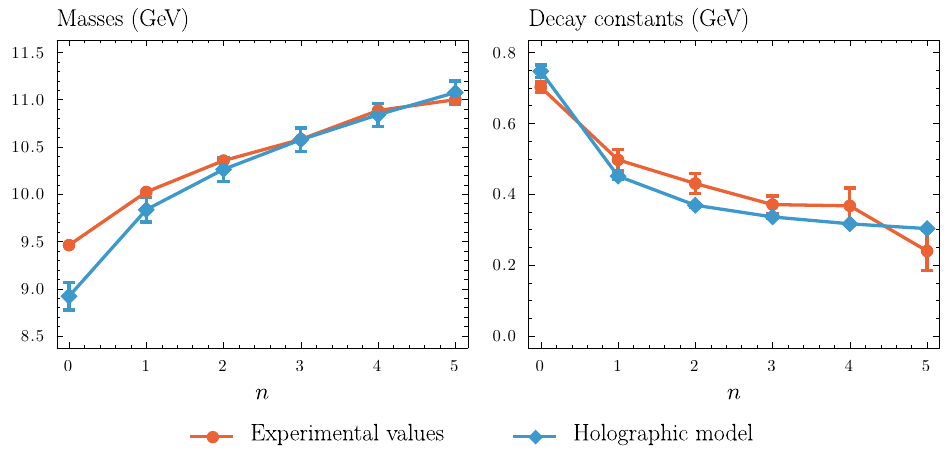}
    \caption{Comparison of bottomonium experimental masses and decay constants with the ones given by the best fit of the holographic model.}
    \label{fig: bottomonium masses and decay constants}
\end{figure}

With each of the 2000 triads of parameters $(κ,M,B)$ found in the fittings described in Sec.~\ref{subsec: Numerical Results - Best Fit Parameters}, we calculated the values of masses and decay constants predicted by the model. Again, we calculated the average predicted values with respective uncertainties from the average and standard deviation of the generated distributions. The results are shown in Tables~\ref{tab: charmonium masses and decay constants} and \ref{tab: bottomonium masses and decay constants}, as well as in Figs.~\ref{fig: charmonium masses and decay constants} and \ref{fig: bottomonium masses and decay constants}.

\pagebreak

\begin{table}[H]
\centering
\begin{tabular}{ccccc}
    \rule{3.8em}{0pt} & \rule{9.0em}{0pt} & \rule{8.0em}{0pt} & \rule{8.0em}{0pt} & \rule{8.0em}{0pt} \\\hline\hline
    \multicolumn{5}{c}{\xrowht{3.5ex} Charmonium masses and decay constants} \\\hline\hline\xrowht{8.5ex}
    State &
    \makecell{Experimental\\masses\\$(\si{\giga\electronvolt})$} &
    \makecell{Predicted\\masses\\$(\si{\giga\electronvolt})$} &
    \makecell{Experimental\\decay constants\\$(\si{\mega\electronvolt})$} &
    \makecell{Predicted\\decay constants\\$(\si{\mega\electronvolt})$}
    \\\hline\hline
    $1S$  & $ 3.09690(0 \pm  6) $ & $3.210 \pm 0.053$ & $ 415 \pm  4 $ & $401 \pm 11$ \\\hline
    $2S$  & $ 3.6860(97 \pm 10) $ & $3.744 \pm 0.048$ & $ 290 \pm  9 $ & $260 \pm  5$ \\\hline
    $3S$  & $    4.04(0 \pm  4) $ & $4.037 \pm 0.046$ & $ 187 \pm 21 $ & $221 \pm  3$ \\\hline
    $4S$  & $    4.41(5 \pm  5) $ & $4.262 \pm 0.045$ & $ 125 \pm 25 $ & $202 \pm  3$ \\\hline
    $5S$  & $    4.6(23 \pm 10) $ & $4.454 \pm 0.045$ & $     -      $ & $    -     $  \\\hline\hline
\end{tabular}
\caption{Comparison of charmonium experimental masses and decay constants with the ones given by the best fit of the holographic model.}
\label{tab: charmonium masses and decay constants}
\end{table}

\vspace*{\fill}

\begin{table}[H]
\centering
\vspace{-\baselineskip}
\begin{tabular}{ccccc}
    \rule{3.8em}{0pt} & \rule{9.0em}{0pt} & \rule{8.0em}{0pt} & \rule{8.0em}{0pt} & \rule{8.0em}{0pt} \\\hline\hline
    \multicolumn{5}{c}{\xrowht{3.5ex} Bottomonium nasses and decay constants}  \\\hline\hline\xrowht{8.5ex}
    State &
    \makecell{Experimental\\masses\\$(\si{\giga\electronvolt})$} &
    \makecell{Predicted\\masses\\$(\si{\giga\electronvolt})$} &
    \makecell{Experimental\\decay constants\\$(\si{\mega\electronvolt})$} &
    \makecell{Predicted\\decay constants\\$(\si{\mega\electronvolt})$}
    \\\hline\hline
    $1S$  & $  9.460(4 \pm 1) $                 & $  8.92 \pm 0.14 $ & $ 702 \pm 14 $ & $ 747 \pm 18 $ \\\hline
    $2S$  & $ 10.023(4 \pm 5) $                 & $  9.84 \pm 0.13 $ & $ 497 \pm 29 $ & $ 451 \pm  8 $ \\\hline
    $3S$  & $ 10.355(1 \pm 5) $                 & $ 10.26 \pm 0.13 $ & $ 430 \pm 28 $ & $ 369 \pm  6 $ \\\hline
    $4S$  & $ 10.57(94 \pm 12) $                & $ 10.57 \pm 0.12 $ & $ 371 \pm 24 $ & $ 336 \pm  5 $ \\\hline
    $5S$  & $ 10.88(52_{\;-\;16}^{\;+\;26}) $ & $ 10.84 \pm 0.12 $ & $ 370 \pm 50 $ & $ 316 \pm  4 $ \\\hline
    $6S$  & $ 11.00(0 \pm 4)   $                & $ 11.07 \pm 0.12 $ & $ 240 \pm 60 $ & $ 303 \pm  4 $  \\\hline\hline
\end{tabular}
\caption{Comparison of bottomonium experimental masses and decay constants with the ones given by the best fit of the holographic model.}
\label{tab: bottomonium masses and decay constants}
\end{table}

\vspace*{\fill}

\pagebreak

\vspace*{\fill}

\vspace*{\fill}

\pagebreak

Using Eqs.~\eqref{eq: parameter interpretation - 1} and \eqref{eq: parameter interpretation - 2} and the values of the parameters $κ$ and $M_q$, we find the charm and bottom quark masses and the string tensions for each meson in this case. From charmonium parameters, we have
\begin{gather}
    m_c = \frac{M_c}{2} = \SI{1.55 \pm 0.03}{\GeV}
    \qquad\text{and}\qquad
    σ_c = \frac{κ_c^3}{4 M_c} = \SI{0.125 \pm .004}{\GeV}^2,
\end{gather}
while for bottomonium we obtain
\begin{gather}
    m_b = \frac{M_b}{2} = \SI{4.62 \pm 0.20}{\GeV}
    \qquad\text{and}\qquad
    σ_b = \frac{κ_b^3}{4 M_b} = \SI{0.254 \pm 0.025}{\GeV}^2,
\end{gather}
which are of the same order of magnitude of the expected values. In \cite{ParticleDataGroup:2020ssz} one finds that charm and bottom masses in the $\overline{\mathrm{MS}}$ scheme are $m_c = \SI{1.27}{\GeV}$ and $m_b = \SI{4.18}{\GeV}$, while the string tension is found to be about $σ = \SI{0.2}{\GeV^2}$ in models based on the Cornell potential for both charmonium and bottomonium.


\vspace{.5\baselineskip}
\subsection{Dilatons and potentials}
\label{subsec: Numerical Results - Dilatons and potentials}

\begin{figure}[htb]
    \vspace{\baselineskip}
    \includegraphics{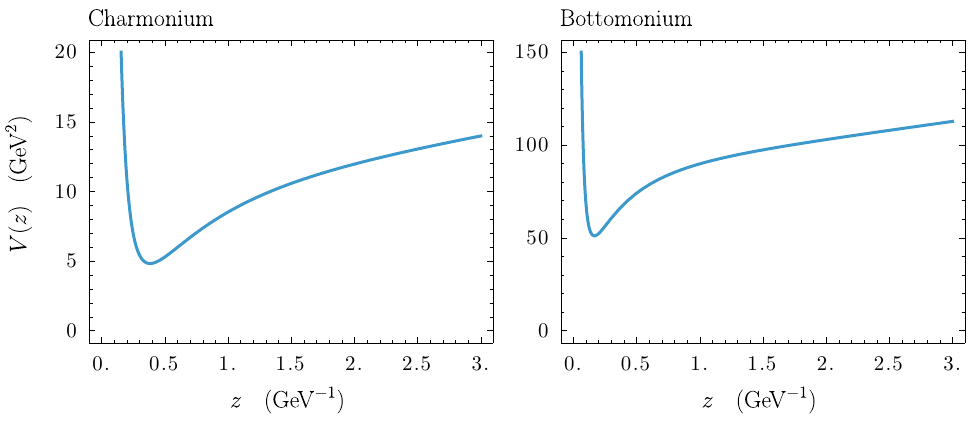}
    \caption{Schrödinger-like potentials $V$ of Eq.~\eqref{eq: Shrodinger-like potential - final model} for the model with the parameter values presented in Table~\ref{tab: best fits}.}
    \label{fig: potentials}
\end{figure}

\begin{figure}[htb]
    \centering
    \includegraphics{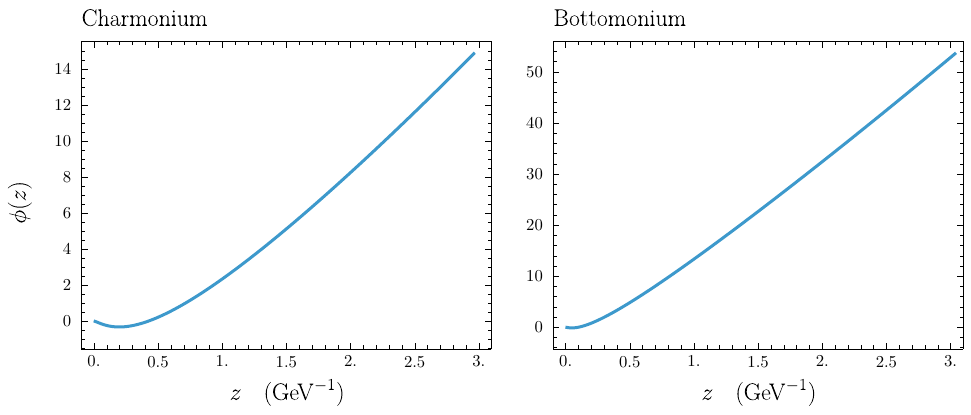}
    \caption{Dilatons for the model with the parameter values presented in Table~\ref{tab: best fits}.}
    \label{fig: dilatons}
\end{figure}

In Figs.~\ref{fig: potentials} and \ref{fig: dilatons} we plot the potentials $V$ of Eq.~\eqref{eq: Shrodinger-like potential - final model} for charmonium and bottomonium and corresponding dilatons with the parameters of Table~\ref{tab: best fits} that provide the best fits.

\vspace{.5\baselineskip}
\section{Conclusions}
\label{sec: Conclusions}

In this work, we developed a holographic model for heavy vector mesons that incorporates features that are present in the phenomenology of quarkonia states. Starting from the soft-wall model, we discussed in Sec.~III how to introduce a mass shift to account for the constituent quark masses. Then, in Sec.~IV we presented a procedure that makes it possible to find a Regge trajectory increasing with a factor $n^{2/3}$ for large $n$ radial excitations. This is achieved by investigating, via the WKB approximation, what kind of term in the potential leads to such a behavior. 

Then we investigated the changes that should be made on the potential in order to reproduce the phenomenological expression of Eq.~\eqref{eq: phenomenological Regge trajectory - 1}. Finally, in order to fit the decay constants, that depend on the behavior of the potential at small $z$, we included a term proportional to $ ē^{-κz}/\sqrt{κz} $.
 
The results obtained in this work represent significant improvements over previous models. In particular, there are two differences with respect to previous holographic works. One is the fact that we used the Langer correction \eqref{eq: Langer transformation} in order to find the proper form of the Regge trajectories; the other is the fact that we found a way to fit the decay constants without spoiling the mass fits.

For the masses and decay constants of charmonium states we obtained a combined error CNRMSE of $14.8\%$. For the bottomnium states the CNRMSE obtained is $11.9\%$. The parameters $M_q$ and $κ$ have interesting physical interpretations as the sum of constituent quark masses and a quantity related to the string tension, respectively, with values comparable to those from potential models.

Future extensions could explore the thermal dissociation of quarkonia in a quark-gluon plasma, incorporating finite temperature, density, magnetic fields, or angular momentum.

Overall, the model presented here provides a bottom-up AdS/QCD framework for studying heavy mesons, representing a bridge between the QSSE \cite{Chen:2018hnx} approach and gauge/gravity duality.


\appendix


\vspace{2\baselineskip}

\noindent {\bf Acknowledgments:} The authors are partially supported by CNPq --- Conselho Nacional de Desenvolvimento Científico e Tecnológico, by FAPERJ --- Fundação Carlos Chagas Filho de Amparo à Pesquisa do Estado do Rio de Janeiro, and by Coordenação de Aperfeiçoamento de Pessoal de Nível Superior --- Brasil (CAPES), Finance Code 001.

\vspace{.5\baselineskip}


\setstretch{1.19}

\bibliography{bibliography}


\end{document}

%% file: configs.tex
\usepackage[english]{babel}
\usepackage{amsmath, amssymb, bm}
\usepackage{amsfonts}  
\usepackage{mathtools} 
\usepackage[normalem]{ulem} 

\usepackage[
    format = hang,
    justification = centering,
    labelfont = bf,
]{caption}
\usepackage{stackengine}
\newcommand\xrowht[2][0]{\addstackgap[.5\dimexpr#2\relax]{\vphantom{#1}}}

\usepackage{float}
\usepackage{graphicx, subcaption} 
\usepackage{tabularx, makecell}
\usepackage{array}
\usepackage{setspace} 

\usepackage{ifthen} 
\usepackage{calc}
\usepackage[ligature]{semantic} 
\usepackage{siunitx} 
\usepackage{csquotes}
\MakeOuterQuote{"}

\makeatletter
\AtBeginDocument{
    \check@mathfonts
    \fontdimen13\textfont2=4.5pt 
    \fontdimen14\textfont2=4.5pt 
    \fontdimen15\textfont2=4.2pt 
    \fontdimen16\textfont2=2.8pt 
    \fontdimen17\textfont2=2.8pt 
}
\makeatother

\let\originalleft\left
\let\originalright\right
\renewcommand{\left}{\mathopen{}\mathclose\bgroup\originalleft}
\renewcommand{\right}{\aftergroup\egroup\originalright}

\DeclareSymbolFont{mylargesymbols}{OMX}{ccex}{m}{n}
\SetSymbolFont{mylargesymbols}{bold}{OMX}{cmex}{m}{n}
\DeclareMathDelimiter{\lbrace}{\mathopen}{symbols}{"66}{mylargesymbols}{"08}
\DeclareMathDelimiter{\rbrace}{\mathclose}{symbols}{"67}{mylargesymbols}{"09}
\DeclareMathDelimiter{(}{\mathopen}{operators}{"28}{mylargesymbols}{"00}
\DeclareMathDelimiter{)}{\mathclose}{operators}{"29}{mylargesymbols}{"01}
\DeclareMathDelimiter{[}{\mathopen}{operators}{"5B}{mylargesymbols}{"02}
\DeclareMathDelimiter{]}{\mathclose}{operators}{"5D}{mylargesymbols}{"03}
\DeclareMathSymbol{\braceld}{\mathord}{mylargesymbols}{"7A}
\DeclareMathSymbol{\bracerd}{\mathord}{mylargesymbols}{"7B}
\DeclareMathSymbol{\bracelu}{\mathord}{mylargesymbols}{"7C}
\DeclareMathSymbol{\braceru}{\mathord}{mylargesymbols}{"7D}

\usepackage{color}
\usepackage{transparent}	

\definecolor{darkgreen}{RGB}{0,100,0}

\newcommand{\AdS}{\mathrm{AdS}}
\newcommand{\SW}{\mathrm{SW}}

\newcommand{\romanII}{I\!I}
\newcommand{\romanIII}{I\!I\!I}
\newcommand{\romanIV}{IV}

\newcommand{\sqrtsum}[2]{\sum\!\raisebox{-0.5ex}{$\rule{0ex}{2.3ex}_{\,#1}^{#2}$}}
\newcommand{\tssum}{\textstyle\sum}

\DeclareMathOperator{\Ai}{Ai}
\DeclareMathOperator{\Bi}{Bi}
\DeclareMathOperator{\Beta}{B}
\DeclareMathOperator{\mean}{mean}

\newcommand{\bvec}[1]{\ensuremath{\bm{#1}}}
\let\mgt\gg 
\let\mlt\ll 

\renewcommand{\gg}{\bvec{g}}

\renewcommand{\ll}{\bvec{\ell}}

\newcommand  {\xx}{\bvec{x}}

\renewcommand{\cal}[1]{\ensuremath{\mathcal{#1}}}

\newcommand{\calO}{\cal{O}}

\newcommand{\PAR}[2][-1]{
    \ifthenelse
    { \equal{#1}{-1} }
    { \ensuremath{ {\left( #2 \right) } } }
    { \ensuremath{
            {               \left( \rule{0pt}{ 0.75ex * #1 + 0.50ex } \right. \hspace{-1pt} }
            #2
            { \hspace{-1pt} \left. \rule{0pt}{ 0.75ex * #1 + 0.50ex } \right)               }
} } }

\newcommand{\COL}[2][-1]{
    \ifthenelse
    { \equal{#1}{-1} }
    { \ensuremath{ {\left[ #2 \right] } } }
    { \ensuremath{
            {               \left[ \rule{0pt}{ 0.75ex * #1 + 0.50ex } \right. \hspace{-1pt} }
            #2
            { \hspace{-1pt} \left. \rule{0pt}{ 0.75ex * #1 + 0.50ex } \right]               }
} } }

\newcommand{\CHA}[2][-1]{
    \ifthenelse
    { \equal{#1}{-1} }
    { \ensuremath{ {\left\{ #2 \right\} } } }
    { \ensuremath{
            {              \left\{ \rule{0pt}{ 0.75ex * #1 + 0.50ex} \right.  \hspace{-1pt} }
            #2
            { \hspace{-1pt} \left. \rule{0pt}at{ 0.75ex * #1 + 0.50ex} \right\}               }
} } }

\newcommand{\LCHA}[1][-1]{%
    \ensuremath{ \hspace{-1pt} \left\{ \rule{0pt}{ 0.75ex * #1 + 0.50ex} \right.  \hspace{-1pt} }
}

\newcommand{\at}[2][-1]{
    \ifthenelse
    { \equal{#1}{-1} }
    { \ensuremath{ {\left.\, #2 \right| } } }
    { \ensuremath{
            {           \!\!\left. \rule{0pt}{ 0.75ex * #1 + 0.50ex} \right. \hspace{-1pt} }
            \,#2
            { \hspace{-1pt} \left. \rule{0pt}{ 0.75ex * #1 + 0.50ex} \right|              }
} } }

\newcommand{\abs}[2][-1]{
    \ifthenelse
    { \equal{#1}{-1} }
    { \ensuremath{ {\left| #2 \right| } } }
    { \ensuremath{
            {               \left| \rule{0pt}{ 0.75ex * #1 + 0.50ex} \right. \hspace{-1pt} }
            #2
            { \hspace{-1pt} \left. \rule{0pt}{ 0.75ex * #1 + 0.50ex} \right|               }
} } }

\newcommand{\average}[2][0]{
    \ifthenelse
    { \equal{#1}{-1} }
    { \ensuremath{ {\left< #2 \right> } } }
    { \ensuremath{
        {               \left< \rule{0pt}{ 0.75ex * #1 + 0.50ex} \right. \hspace{-1pt} }
        #2
        { \hspace{-1pt} \left. \rule{0pt}{ 0.75ex * #1 + 0.50ex} \right>               }
} } }

\newcommand{\omatrix}[1]{\begin{matrix}#1\end{matrix}}

\newcommand{\complexi}{\hspace{.15ex}\textrm{i}\hspace{.15ex}}
\newcommand{\exponentiale}{\hspace{.15ex}\textrm{e}\hspace{.15ex}}

\makeatletter

\@namedef{u8:\detokenize{α}}{\alpha}
\@namedef{u8:\detokenize{β}}{\beta}
\@namedef{u8:\detokenize{γ}}{\gamma}
\@namedef{u8:\detokenize{δ}}{\delta}
\@namedef{u8:\detokenize{ϵ}}{\epsilon}
\@namedef{u8:\detokenize{ε}}{\varepsilon}
\@namedef{u8:\detokenize{ζ}}{\zeta}
\@namedef{u8:\detokenize{η}}{\eta}
\@namedef{u8:\detokenize{θ}}{\theta}
\@namedef{u8:\detokenize{ϑ}}{\vartheta}
\@namedef{u8:\detokenize{κ}}{\kappa}
\@namedef{u8:\detokenize{λ}}{\lambda}
\@namedef{u8:\detokenize{μ}}{\mu}
\@namedef{u8:\detokenize{ν}}{\nu}
\@namedef{u8:\detokenize{ξ}}{\xi}
\@namedef{u8:\detokenize{π}}{\pi}
\@namedef{u8:\detokenize{ρ}}{\rho}
\@namedef{u8:\detokenize{ϱ}}{\varrho}
\@namedef{u8:\detokenize{σ}}{\sigma}
\@namedef{u8:\detokenize{ς}}{\varsigma}
\@namedef{u8:\detokenize{τ}}{\tau}
\@namedef{u8:\detokenize{υ}}{\upsilon}
\@namedef{u8:\detokenize{ϕ}}{\phi}
\@namedef{u8:\detokenize{φ}}{\varphi}
\@namedef{u8:\detokenize{χ}}{\chi}
\@namedef{u8:\detokenize{ψ}}{\psi}
\@namedef{u8:\detokenize{ω}}{\omega}
\@namedef{u8:\detokenize{ϖ}}{\varpi}

\@namedef{u8:\detokenize{Γ}}{\Gamma}
\@namedef{u8:\detokenize{Δ}}{\Delta}
\@namedef{u8:\detokenize{Θ}}{\Theta}
\@namedef{u8:\detokenize{Λ}}{\Lambda}
\@namedef{u8:\detokenize{Ξ}}{\Xi}
\@namedef{u8:\detokenize{Π}}{\Pi}
\@namedef{u8:\detokenize{Σ}}{\Sigma}
\@namedef{u8:\detokenize{ϒ}}{\Upsilon}
\@namedef{u8:\detokenize{Φ}}{\Phi}
\@namedef{u8:\detokenize{Ψ}}{\Psi}
\@namedef{u8:\detokenize{Ω}}{\Omega}

\@namedef{u8:\detokenize{±}}{\pm}
\@namedef{u8:\detokenize{²}}{^{2}}
\@namedef{u8:\detokenize{³}}{^{3}}
\@namedef{u8:\detokenize{¹}}{^{1}}
\@namedef{u8:\detokenize{⁰}}{^{0}}
\@namedef{u8:\detokenize{·}}{\hspace{-0.05ex}\cdot\hspace{-0.05ex}}
\@namedef{u8:\detokenize{×}}{\hspace{-0.05ex}\times\hspace{-0.05ex}}
\@namedef{u8:\detokenize{î}}{\complexi}
\@namedef{u8:\detokenize{ĩ}}{\complexi}
\@namedef{u8:\detokenize{ħ}}{\hbar}
\@namedef{u8:\detokenize{ḏ}}{\mathrm{d}}
\@namedef{u8:\detokenize{ē}}{\exponentiale}
\@namedef{u8:\detokenize{†}}{\dagger}
\@namedef{u8:\detokenize{‡}}{\ddagger}
\@namedef{u8:\detokenize{ı}}{\imath}
\@namedef{u8:\detokenize{ℓ}}{\ell}
\@namedef{u8:\detokenize{←}}{\leftarrow}
\@namedef{u8:\detokenize{→}}{\rightarrow}
\@namedef{u8:\detokenize{∂}}{\partial}
\@namedef{u8:\detokenize{≠}}{\neq}
\@namedef{u8:\detokenize{≤}}{\leq}
\@namedef{u8:\detokenize{≥}}{\geq}
\@namedef{u8:\detokenize{∞}}{\infty}
\@namedef{u8:\detokenize{∫}}{\int}
\@namedef{u8:\detokenize{≈}}{\approx}
\@namedef{u8:\detokenize{≡}}{\equiv}
\@namedef{u8:\detokenize{□}}{\Box}
\@namedef{u8:\detokenize{ⵈ}}{\cdots}
\@namedef{u8:\detokenize{ø}}{\phantom{0}}

\makeatother

\expandafter\newcommand\csname u8:\detokenize{√}\endcsname[2][]{%
    \ifthenelse
    { \equal{#1}{} }
    { \sqrt{#2} }
    { \sqrt[#1]{#2} }
}

\mathlig{<<}{\mlt}
\mathlig{>>}{\mgt}
\mathlig{!=}{\neq}
\mathlig{=!}{\neq}
\mathlig{/=}{\neq}
\mathlig{=/}{\neq}
\mathlig{<>}{\neq}
\mathlig{><}{\neq}
\mathlig{==}{\equiv}
\mathlig{='}{\simeq}
\mathlig{<=}{\leq}
\mathlig{=<}{\leq}
\mathlig{>=}{\geq}
\mathlig{=>}{\geq}

\mathlig{+-}{\pm}
\mathlig{-+}{\mp}
\mathlig{-->}{\rightarrow}
\mathlig{--->}{\longrightarrow}
\mathlig{<--}{\leftarrow}
\mathlig{<---}{\longleftarrow}
\mathlig{<->}{\leftrightarrow}
\mathlig{<-->}{\longleftrightarrow}
\mathlig{|->}{\mapsto}
\mathlig{|-->}{\longmapsto}
\mathlig{==>}{\Rightarrow}
\mathlig{===>}{\Longrightarrow}
\mathlig{==>}{\Leftarrow}
\mathlig{===>}{\Longleftarrow}
\mathlig{<=>}{\leftrightarrow}
\mathlig{<==>}{\longleftrightarrow}

